\documentclass{elsart}

\usepackage{color}

 \usepackage{graphicx}

\usepackage{amssymb}
\usepackage{multirow}

\usepackage{color}

\begin{document}

\begin{frontmatter}



\title{Analysis of the impact degree distribution in metabolic networks
using branching process approximation}


\author[JST,Tokyo]{Kazuhiro Takemoto}
\author[BIC]{Takeyuki Tamura}
\author[HongKong]{Yang Cong}
\author[HongKong]{Wai-Ki Ching}
\author[Paris,Curie,Inserm]{Jean-Philippe Vert}
\author[BIC,corr]{Tatsuya Akutsu}
\ead{takutsu@kuicr.kyoto-u.ac.jp}

\address[JST]{PRESTO, Japan Science and Technology Agency, Kawaguchi, Saitama 332-0012, Japan}

\address[Tokyo]{Department of Biophysics and Biochemistry, University of Tokyo, Hongo 7-3-1, Bunkyo-ku, Tokyo 113-0033, Japan}

\address[BIC]{Bioinformatics Center, Institute for Chemical Research, Kyoto University, Gokasho, Uji, Kyoto, 611-0011, Japan}

\address[HongKong]{Department of Mathematics, The University of Hong Kong, Pokfulam Road, Hong Kong}

\address[Paris]{Centre for Computational Biology, Mines ParisTech, 35 rue Saint-Honor{\'e}, 77305 Fontainebleau cedex, France}

\address[Curie]{Institut Curie, 75248 Paris, France}

\address[Inserm]{INSERM, U900, 75248 Paris, France}

\corauth[corr]{Corresponding author.}

\begin{abstract}
Theoretical frameworks to estimate the tolerance of metabolic networks to various failures are important to evaluate the robustness of biological complex systems in systems biology.
In this paper, we focus on a measure for robustness in metabolic networks, namely, the {\it impact degree}, and propose an approximation method to predict the probability distribution of impact degrees from metabolic network structures using the theory of branching process.
We demonstrate the relevance of this method by testing it on real-world metabolic networks.
Although the approximation method possesses a few limitations, it may be a powerful tool for evaluating metabolic robustness.
\end{abstract}


\end{frontmatter}

\section{Introduction}
Robustness is a key feature in the analysis of complex systems, especially for complex biological systems.
Many organisms have strong adaptability to environmental changes or failures in some of their components, and can live even if some of their genes are mutated.
In particular, it is known that cancer cells are very robust \cite{kitano04}.
Therefore, understanding the origin of robustness of living cells has become an important research topic.

In particular, extensive studies have focused on the analysis of structural robustness of metabolic networks.
Structural robustness refers to the tolerance of the system's behavior to changes in the structure of networks, and most existing studies focus
on changes caused by knockout of gene(s) or enzyme(s). 
One of the reasons why extensive studies have been done on structural robustness of metabolic networks is that rather accurate data of metabolic networks are available via such databases as the Kyoto Encyclopedia of Genes and Genomes (KEGG) \cite{kanehisa10} and the Encyclopedia of {\it Escherichia coli} K-12 Genes and Metabolism (EcoCyc) \cite{keseler11}, and kinetic parameters, which are not necessarily available, are not required.

In order to analyze the structural robustness of metabolic networks, the \emph{flux balance analysis} (FBA) methods have been widely used.
In many of these approaches, \emph{elementary flux modes} (EFMs) play a key role, where an EFM is a minimal set of reactions that can operate at steady state \cite{papin04}.
Based on FBA and/or EFM, several studies have focused on finding a minimum reaction cut \cite{acuna09,burgard03,haus08,klamt04}, that is, a minimum set of reaction (or enzyme) removals which prevent the production of a specified set of compounds.
Other FBA-based measures of robustness have also been proposed.
Behre et al. proposed a measure based on the number of remaining EFMs after knockout versus the number of EFMs in the unperturbed situation \cite{behre08}.
Deutscher et al. proposed another measure using the Shaply value from game theory \cite{deutscher08}.

Other approaches have been proposed based on Boolean models of metabolic networks in which reactions and compounds are modeled as AND and OR nodes, respectively.
Handorf et al. analyzed robustness of metabolic networks by introducing the concept of {\em scope} \cite{handorf05}.
Li et al., Sridhar et al., and Tamura et al. developed integer programming-based methods for finding a minimum reaction cut in Boolean models of metabolic networks \cite{Lemke2004,li09,sridhar08,tamura09}.
Smart et al. defined the {\em topological flux balance} (TFB) criterion based on a Boolean model of metabolic networks and analyzed the {\em damage} (number of reactions) caused by knockout of a single reaction under TFB \cite{smart08}.
Jiang et al. defined and analyzed the \emph{impact degree}, which is the number of reactions inactivated by knockout of a specific reaction \cite{jiang09}.
Although there are some differences in the treatment of reversible reactions, the damage and the impact degree are very similar concepts.
Cong et al. extended the impact degree for knockout of multiple reactions \cite{cong09}.

In this paper, we study the distribution of the impact degree caused by random knockout of a single reaction using the theory of branching process \cite{lee04,saichev05}.
In order to analyze earthquakes, Saichev et al. proposed a branching process with power-law distributions of offspring $d$: $P(d) \propto 1/d^{\gamma+1}$, where $\gamma$ is some constant, and approximately derived the distribution of the total number of offsprings \cite{saichev05}.
We regard propagation of the impact of knockout of a reaction as a branching process, and apply their method to estimate the impact degree distribution, where the impact degree in our problem corresponds to the total number of offsprings in the branching process.
In order to apply this method, we develop a simple method for estimating the offspring distribution in a metabolic network.
Although Smart et al. have already applied percolation theory and branching process to analysis of the size distribution of rigid clusters, defined as clusters of contagion nodes that do not contain any branched metabolite nodes (see Fig. 3A in \cite{smart08}), they did not explicitly estimate the damage distribution (i.e., the impact degree distribution).
We finally show an estimation method for the damage distributions.
The proposed method is applied to analysis of metabolic networks of
four species: \emph{Escherichia coli}, \emph{Bacillus subtilis},
\emph{Saccharomyces cerevisiae} and \emph{Homo sapiens}. 
The results show good agreement of impact degree distributions between empirical results and theoretical estimates.

\section{Impact degree}
\label{sec:imp_deg}

Jiang et al. proposed the \emph{impact degree} as a measure of the importance
of each reaction in a metabolic network \cite{jiang09}.
The impact degree is defined as the number of inactivated reactions caused
by knockout of a single reaction.
However, they did not consider the effect of cycles in metabolic networks.
Since cycles play an important role in metabolic networks,
Cong et al. extended the impact degree so that the effect of cycles is
taken into account \cite{cong09} by using a concept of the 
maximal valid assignment \cite{tamura09}.
Here, we briefly review the definition of this extended impact degree \cite{cong09}.

As in other works, we regard each metabolic network as a bipartite directed
graph.
Let $V_c = \{C_1,\ldots,C_m\}$ and $V_r = \{R_1,\ldots,R_n\}$ be
a set of \emph{compound nodes} and a set of \emph{reaction nodes} respectively,
where $V_c \cap V_r = \{\}$.
A \emph{metabolic network} is defined as a directed graph $G(V_c \cup V_r,E)$
in which either $(u \in V_c) \land (v \in V_r)$ or $(u \in V_r) \land (v \in V_c)$
holds for each edge $(u,v) \in E$.\footnote{$A \land B$ means logical AND of $A$
and $B$.}
Each reaction and compound takes one of two states: 0 or 1, where 0 and 1
correspond to inactive and active reactions (compounds), respectively.
Reverse reactions are treated as two irreversible reactions.

In order to define the impact degree of a reaction, we proceed as follows. Suppose that reaction $R_i$ is knocked out.
Then, we start with the global state where all compounds are active
(i.e., $C_k=1$ for all $C_k \in V_c$) and all reactions but $R_i$ are active
(i.e., $R_j=1$ for all $R_j \in V_r \backslash \{ R_i \}$ and $R_i=0$).
Then, we alternatively update the states of reactions and compounds by the following rules.
\begin{enumerate}
\item For each reaction, there are three different compounds: consumed compounds
(i.e., substrates), produced compounds (i.e., products), and
directly unrelated compounds.
\item A reaction is inactivated if any of its consumed or produced
compound is inactivated.
\item  For each compound, there are three different reactions: consuming reactions,
producing reactions, and directly unrelated reaction.
\item A compound is inactivated if all its consuming reactions or all
its producing reactions are inactivated.
\end{enumerate}
Since no activation is possible in this process, the procedure converges to a stable state in a finite number of iterations.
The impact degree of reaction $R_i$ is defined as the number of inactivated reactions in the stable state.
This procedure simultaneously gives the definition of the impact degree
and an algorithm to compute it.

\begin{figure}[th]
\begin{center}
\includegraphics{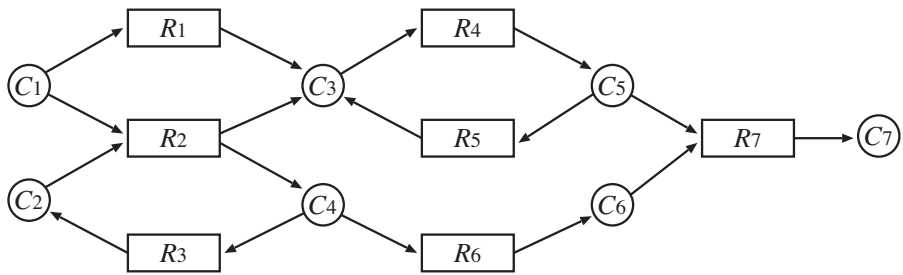}
\caption{Example of a metabolic network. Boxes and circles correspond to
reactions and compounds, respectively.}
\label{exnet}
\end{center}
\end{figure}

Let us illustrate the above process with the metabolic network
shown in Fig.~\ref{exnet}.
Suppose that reaction $R_2$ is knocked out.
Then, the states of nodes change as shown in Table~\ref{exknock1}.
Since four reactions (including $R_2$) are inactivated in the stable state,
the resulting impact degree is four.
Next, suppose that reaction $R_6$ is knocked out.
In this case, the states of nodes change as shown in Table~\ref{exknock2}
and the resulting impact degree is two.
It is to be noted that $C_4$ is not inactivated because
$R_2$ is still active in this case.

\begin{table}[ht]
\begin{center}
\caption{Impact degree calculation when $R_2$ in Fig.~\ref{exnet} is knocked out.}
\label{exknock1}

\medskip

\begin{tabular}{ccccccc|ccccccc}
\hline
$R_1$ & $R_2$ & $R_3$ & $R_4$ & $R_5$ & $R_6$ & $R_7$ &
$C_1$ & $C_2$ & $C_3$ & $C_4$ & $C_5$ & $C_6$ & $C_7$ \\
\hline
1 & 0 & 1 & 1 & 1 & 1 & 1 & 1 & 1 & 1 & 1 & 1 & 1 & 1 \\
1 & 0 & 1 & 1 & 1 & 1 & 1 & 1 & 0 & 1 & 0 & 1 & 1 & 1 \\
1 & 0 & 0 & 1 & 1 & 0 & 1 & 1 & 0 & 1 & 0 & 1 & 1 & 1 \\
1 & 0 & 0 & 1 & 1 & 0 & 1 & 1 & 0 & 1 & 0 & 1 & 0 & 1 \\
1 & 0 & 0 & 1 & 1 & 0 & 0 & 1 & 0 & 1 & 0 & 1 & 0 & 1 \\
1 & 0 & 0 & 1 & 1 & 0 & 0 & 1 & 0 & 1 & 0 & 1 & 0 & 0 \\
\hline
\end{tabular}
\end{center}
\end{table}

\begin{table}[ht]
\begin{center}
\caption{Impact degree calculation when $R_6$ in Fig.~\ref{exnet} is knocked out.}
\label{exknock2}

\medskip

\begin{tabular}{ccccccc|ccccccc}
\hline
$R_1$ & $R_2$ & $R_3$ & $R_4$ & $R_5$ & $R_6$ & $R_7$ &
$C_1$ & $C_2$ & $C_3$ & $C_4$ & $C_5$ & $C_6$ & $C_7$ \\
\hline
1 & 1 & 1 & 1 & 1 & 0 & 1 & 1 & 1 & 1 & 1 & 1 & 1 & 1 \\
1 & 1 & 1 & 1 & 1 & 0 & 1 & 1 & 1 & 1 & 1 & 1 & 0 & 1 \\
1 & 1 & 1 & 1 & 1 & 0 & 0 & 1 & 1 & 1 & 1 & 1 & 0 & 1 \\
1 & 1 & 1 & 1 & 1 & 0 & 0 & 1 & 1 & 1 & 1 & 1 & 0 & 0 \\
\hline
\end{tabular}
\end{center}
\end{table}

\section{Branching process approximation}
\label{sec:branch}
We here explain the branching process approximation for estimating the impact degree distributions of metabolic networks.

The branching process is a stochastic process in which each progenitor generates offsprings according to a fixed probability distribution called the offspring distribution. 
We propose that the branching process approximation is useful for estimating the impact degree distribution because the propagation of an impact on a network is essentially similar to cascading failures, which is a sequence of failures caused by an accident.
The branching process model of cascading failure is a standard Galton-Watson branching process \cite{Harris1989}. The approximation method using branching process has been already applied to loading-dependent cascading failure, and its relevance has been shown \cite{Dobson2005,Kim2010}.
However, the branching process approximation needs the assumption of tree structure of networks; thus, this is a limitation of the branching process approximation because metabolic networks generally have cycles.

\subsection{The number of offsprings in metabolic networks}
To estimate the impact degree distributions using the branching process approximation, we need to define the notion of offsprings for each reaction node in metabolic networks, and the distribution of the number of offsprings.

\begin{figure}[th]
\begin{center}
\includegraphics{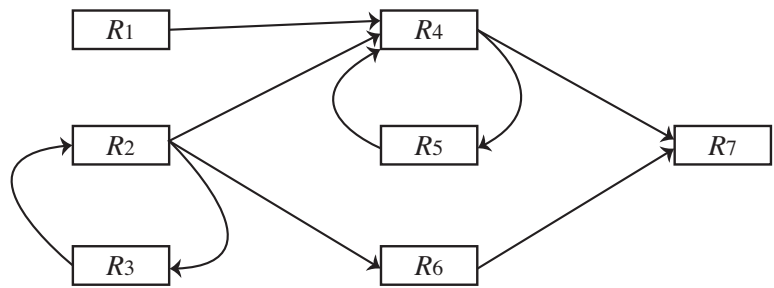}
\caption{The reaction network transformed from the metabolic network in Fig. \ref{exnet}.}
\label{fig:reaction_network}
\end{center}
\end{figure}

For that purpose, we consider the reaction network obtained as the unipartite projection of the metabolic network, where we draw an edge from reaction A to reaction B when at least one product of A is a substrate of B \cite{Wagner2001,Light2005,Pal2005,Vitkup2006}. Fig. \ref{fig:reaction_network} shows the reaction network obtained from the metabolic network of Fig. \ref{exnet} by this procedure.

As an easy example, we consider 2 metabolic reactions, A and B. In this case, the edge is drawn from A to B (i.e., A$\to$B) if at least 1 product of reaction A corresponds to at least 1 substrate of reaction B (e.g., the case of a$\to$A$\to$b$\to$B$\to$c, where a, b, and c are chemical compounds).
Through a similar procedure, we obtain a reaction network from a given metabolic network (see Fig. \ref{fig:reaction_network}).

We now observe that the number of reactions inactivated by the failure of a given reaction, which we want to model in the branching process, does not correspond to its outdegree (i.e., the number of out-going edges from a reaction node) in the reaction network because the spreading of an impact is depressed when there are alternative synthetic pathways.
For example, assuming that the reaction $R_1$ is inactive (or disrupted), the cascading of the impact does not occur because the reaction $R_4$ remains active due to the chemical compound generated through the reaction $R_2$.
In contrast, the cascading of the impact continues when the reaction $R_2$ becomes inactive for instance because the reaction $R_6$ is dependent on this reaction only.
In short, the impact spreads though reactions whose substrates are synthesized via unique metabolic reactions (i.e., reaction nodes with the indegree of 1) when assuming tree structures of networks.

Based on this analysis, we define the number of offsprings for each reaction node in metabolic networks as follows:
\begin{equation}
d_i =
\left\{
\begin{array}{ll}
k^{\mathrm{out}}_i & (\mathrm{if} \ k^{\mathrm{in}}_i=1) \\
0 & (\mathrm{otherwise}) \\
\end{array}
\right.,
\end{equation}
where $k^{\mathrm{out}}_i$ and $k^{\mathrm{in}}_i$ are the outdegree and indegree of reaction node $i$ in the reaction network, respectively.

\subsection{Branching process models}
To analytically estimate the impact degree distributions, it is useful to assume that the number of offsprings for each node follows a probability distribution.
We here consider two types of distributions, which are frequently observed in real worlds.

\subsubsection{Poisson model}
The simplest case of branching processes is a Poisson branching process (hereafter called {\it Poisson model}) in which the number of offsprings $d$ for each progenitor follows the Poisson distribution: $\mu^{d}e^{-\mu}/d !$, where $\mu$ corresponds to the mean of this distribution.
In this model, the total number of offsprings (i.e., impact degree) $r$ is distributed according to the Borel distribution \cite{Dobson2004}:
\begin{equation}
P(r)=(\mu r)^{r-1}\frac{e^{-\mu r}}{r!}.
\label{eq:borel}
\end{equation}

Using Stirling's formula (i.e., $r!\approx \sqrt{2\pi r}r^r e^{-r}$), the above equation leads to the approximation
\begin{equation}
P(r)\propto r^{-3/2}e^{-r(\ln \mu-\mu+1)}.
\label{eq:approx_borel}
\end{equation}
In particular, when $\mu=1$ (i.e., the critical case), the impact degree follows a power-law distribution with exponent $-3/2$.

\subsubsection{Power-law model}
In addition to the Poisson case, we consider the case where the number of offsprings is determined based on a power-law distribution (hereafter called {\it Power-law model}).
Indeed, the number of offsprings for each reaction node is based on the outdegree in metabolic networks.
Since real-world complex networks including metabolic networks have power-law degree distributions \cite{Wagner2001,Jeong2000,Albert2002}, the number of offsprings for each reaction node may also obey a power-law distribution.

Saichev {\it et al.} \cite{saichev05,Saichev2004} showed analytical asymptotic approximations for the distribution $P(r)$ of the total number of offsprings (i.e., impact degree) $r$ in the case where the number of offsprings $d$ for each progenitor follows asymptotically a power-law distribution $P(d) \propto 1/d^{\gamma+1}$ and the mean number of offsprings has a given value $\mu$. In particular they derive the following approximation for large impact degree $r$, in the case $1<\gamma<2$ where the variance of the number of offsprings is infinite:
\begin{equation}
P(r)\simeq\frac{\mu}{\nu r^{1+1/\gamma}}\varphi_\gamma \left( \frac{(1-\mu)r-\mu-1}{\nu r^{1/\gamma}} \right) \,,
\label{eq:P(r)_case3}
\end{equation}
where
\begin{equation}
\varphi_\gamma(x)=\int_0^{\infty} \exp\left[ u^{\gamma} \cos\left(\frac{\pi\gamma}{2}\right)\right] \cos\left[u^\gamma \sin\left(\frac{\pi\gamma}{2}\right)+ux\right]du\,,
\end{equation}
and $\nu = \mu(\gamma-1) \gamma^{1/\gamma - 1} \Gamma(-\gamma)^{1/\gamma}$, where $\Gamma(x)$ is the Gamma function.

When $\mu=1$ (i.e., the critical case), in particular, the distribution of the total number of offsprings obeys the power-law distribution: $P(r)\propto 1/r^{1+1/\gamma}$.

In addition, $P(r)$ in the case of $\gamma > 2$ is approximately similar to that in the case of Poisson model [i.e., Eq. (\ref{eq:borel})] \cite{saichev05}.

\subsubsection{Empirical model}
The above probability distributions may be unsuitable to approximate real-world offspring distributions.
In addition, we also consider a branching process model using empirical offspring distributions (hereafter called {\it Empirical model}).
This way, we can estimate the distribution $P(r)$ of the total number of offsprings (i.e., impact degree distributions) without the approximation of offspring distributions, although the model is analytically intractable.
Moreover, we can evaluate whether the prediction accuracy of the proposed method is influenced by the approximation of offspring distributions or the fidelity of branching processes.

Let $F(s)$ be the probability generating function of the impact degree $r$ (i.e., the total number of offsprings), the function $F(s)$ satisfies the recursive relation \cite{saichev05,Harris1989}:
\begin{equation}
F(s)=f(sF(s)),
\end{equation}
where $f(s)$ denotes the probability generating function of the number $d$ of offsprings of each node.

Using the Lagrange expansion and the relation of $P(r)=(1/r!)d^rF(s)/ds^r|_{s=0}$, the distribution $P(r)$ (i.e., impact degree distribution) is derived from the above implicit equation as the following explicit equation \cite{saichev05,Harris1989}:
\begin{equation}
P(r)=\left.\frac{1}{r!}\frac{d^{r-1}}{ds^{r-1}}\left[f^r(s)\frac{df(s)}{ds}\right]\right|_{s=0} \ (r>0),
\label{eq:P(r)_emp}
\end{equation}
where $f(s)=\sum_{d=0}^{d_{\max}}P(d)s^d$.
The value $d_{\max}$ indicates the maximum of $d$, and the function $P(d)$ corresponds to the probability density function of empirical $d$ (i.e., empirical offspring distribution). In addition, $P(r)=f(0)$ when $r=0$.

\subsubsection{Parameter extraction}
To apply the Poisson and the power-law models, we need to estimate the model parameters, namely the mean $\mu$ and the exponent $\gamma$, from real metabolic networks.

We estimate the mean of the number of offsprings for each reaction node by the empirical average:
\begin{equation}
\mu=\frac{1}{N}\sum_{i=1}^Nd_i,
\label{eq:mean}
\end{equation}
where $N$ is the total number of reaction nodes in a metabolic network.

We estimate the exponent of a power-law offspring distribution using the maximum likelihood estimation method \cite{Newman2005}: 
\begin{equation}
\gamma=|N^*|\left[\sum_{i\in N^*}\ln\frac{d_i}{d_{\mathrm{min}}}\right]^{-1},
\label{eq:gamma}
\end{equation}
where $N^*$ is the set of reaction nodes with $d_i>0$, and $|N^*|$ indicates the total number of such reaction nodes.
$d_{\mathrm{min}}$ is the minimum of $d_i$ in the set of $N^*$.

\section{Evaluation of the branching process approximation}
We evaluated the above estimation methods for the impact degree distributions on several real metabolic networks.

We selected two bacteria [{\it Escherichia coli} (eco) and {\it Bacillus subtilis} (bsu)] and two eukaryotes [{\it Saccharomyces cerevisiae} (sce) and {\it Homo sapiens} (hsa)] whose metabolic pathways have been well-identified. We downloaded the data of their metabolic networks, represented as bipartite networks as shown in Fig. \ref{exnet}, from the KEGG database (version 0.7.1) \cite{kanehisa10,KEGG}.
The parenthetic three-letter codes correspond to KEGG organism identifiers \cite{KEGGorg}.

Based on the KEGG metabolic network data, the impact degree distributions in the metabolic networks were calculated using the method explained in  Sec. \ref{sec:imp_deg}.
Moreover, we constructed the reaction networks of these species, and obtained the offspring distributions.
Using Eqs. (\ref{eq:mean}) and (\ref{eq:gamma}), the model parameters $\mu$ and $\gamma$ were extracted (see Table \ref{table:parameter}).
All metabolic networks show $1<\gamma<2$, implying that the assumption of power-law model is suitable.

\begin{table}[tbhp]
\caption{Model parameters extracted from real metabolic networks. The character \# indicates ``the number of". The number of reaction nodes corresponds to the number of parents in branching processes.}
\label{table:parameter}
\begin{center}
\begin{tabular}{l|ccc}
\hline
\hline
Species & \#Reaction nodes & Mean $\mu$ & Exponent $\gamma$ \\
\hline
{\it Escherichia coli} & 1,467 & 0.68 & 1.52 \\
{\it Bacillus subtilis} & 1,279 & 0.60 & 1.65 \\
{\it Saccharomyces cerevisiae} & 1,172 &  0.54 &  1.73 \\
{\it Homo sapiens} & 1,982 & 0.58 &  1.50 \\
\hline
\hline
\end{tabular}
\end{center}
\end{table}

\begin{figure}[tbp]
\begin{center}
	\includegraphics{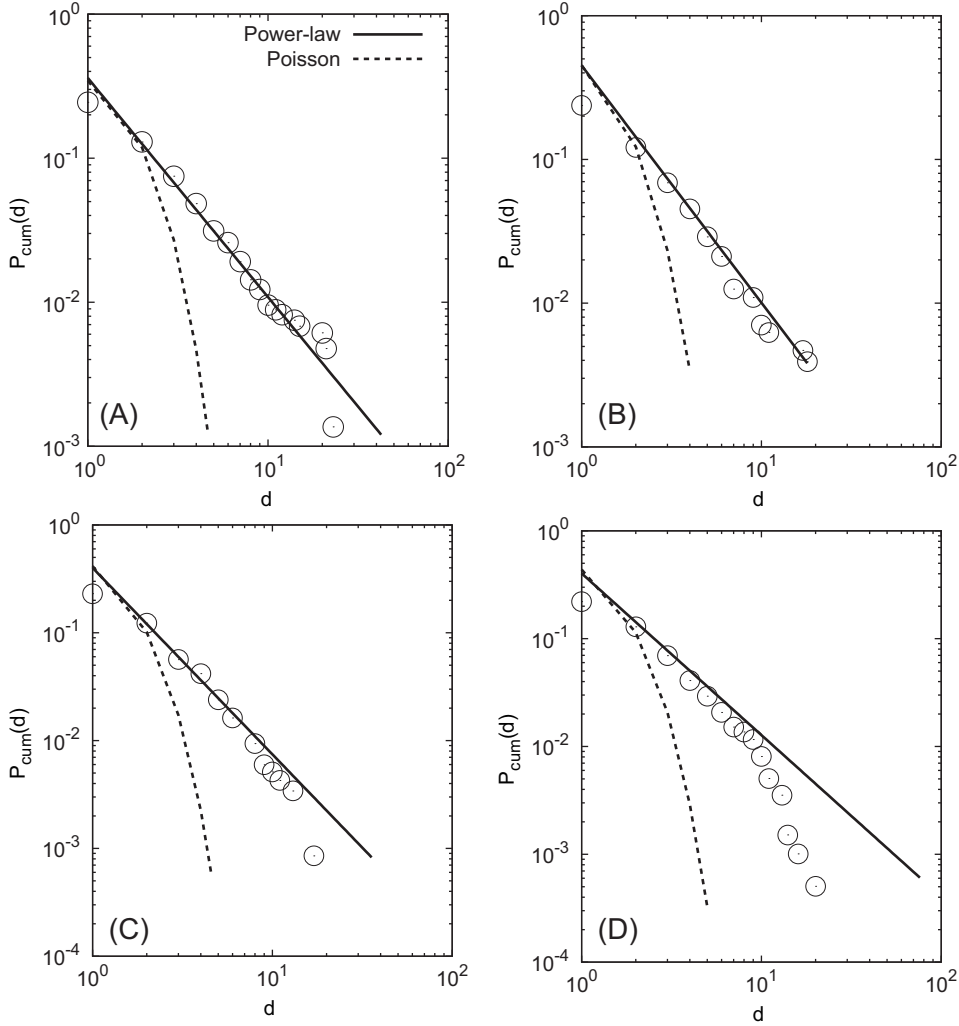}  
	\caption{
	Cumulative offspring distributions of 
	{\it Escherichia coli} (A),
	{\it Bacillus subtilis} (B),
	{\it Saccharomyces cerevisiae} (C), and
	{\it Homo sapiens} (D).
	$P_{cum}(x)$ is defined as $P(X\geq x)$.
	Note that $P_{cum}(1)<1$ because the cumulative distributions also consider the case of $d=0$.
	$P_{cum}(0)$ is not shown due to the logarithmic display.
	The symbols indicate observed data.
	The black solid lines and dashed lines correspond to the cumulative representations of the power-law distribution with the exponent estimated by Eq. (\ref{eq:gamma}) and the Poisson distribution with the mean obtained from Eq. (\ref{eq:mean}), respectively.
	}
	\label{fig:off_dist}
\end{center}
\end{figure}

To test the validity of the offspring distribution models, we compared the fitting results on empirical offspring distributions in real metabolic networks between the power-law distribution and the Poisson distribution.
Fig. \ref{fig:off_dist} shows the cumulative offspring distributions from the real metabolic networks and the cumulative representation of theoretical distributions.
The figure clearly indicates that the power-law distributions are more appropriate for modeling offspring distributions than the Poisson distributions.
However, the power-law distributions may be not the best model because of the poor fittings for the larger $d$ in the case of {\it H. sapiens} (Fig. \ref{fig:off_dist}D)

\begin{figure}[tbp]
\begin{center}
	\includegraphics{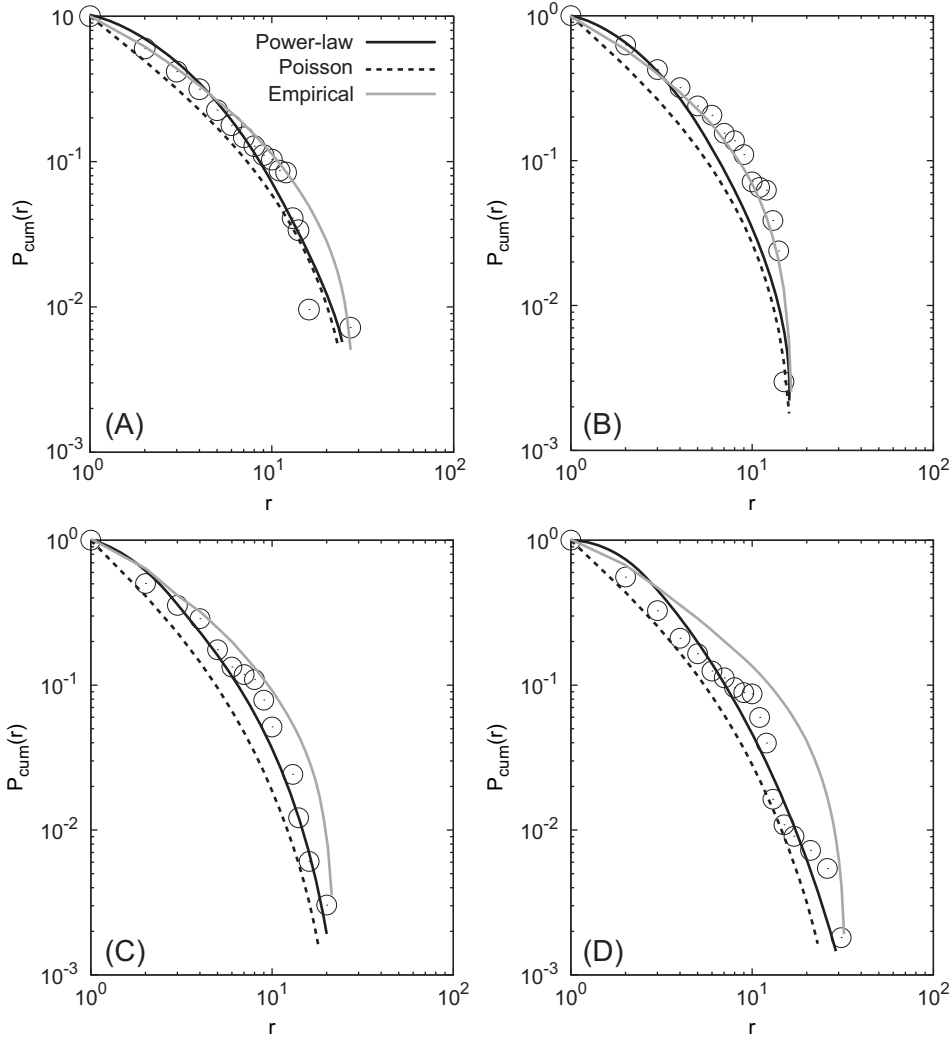}  
	\caption{
	Cumulative impact degree $r$ distributions of 
	{\it Escherichia coli} (A),
	{\it Bacillus subtilis} (B),
	{\it Saccharomyces cerevisiae} (C), and
	{\it Homo sapiens} (D).
	The symbols indicate observed data.
	The black solid lines and dashed lines correspond to the cumulative representations of theoretical distributions of the power-law model and the Poisson model, respectively.
	The gray solid lines are the the cumulative representations of theoretical distributions from the empirical model.
	Note that $P_{cum}(x)$ is defined as $P(X\geq x)$.
	}
	\label{fig:imp_deg_dist}
\end{center}
\end{figure}

Using Eqs. (\ref{eq:borel}), (\ref{eq:P(r)_case3}), and (\ref{eq:P(r)_emp}), we obtained the estimated impact degree distributions using the branching process approximation.
Fig. \ref{fig:imp_deg_dist} shows the comparison between the observed cumulative impact degree distributions and estimated ones.
The theoretical predictions (lines) are in good agreement with the real impact degree distributions (symbols), suggesting the relevance of branching process approximations.
Note that the impact degree distributions does not follow a clear power law and show an exponential cut-off for larger impact degrees because $\mu<1$ (i.e., not the critical case).
Eqs. (\ref{eq:approx_borel}) or (\ref{eq:P(r)_case3}) can explain this distributional tendency.

To evaluate the prediction accuracy for impact degree distributions between the Power-law model and the Poisson model, we evaluated the distributional distance (i.e., Kolmogorov-Smirnov statistics) between the observed distributions and theoretical distributions (Table \ref{table:accuracy}).
The power-law model is better than the Poisson model on all networks. The empirical model outperforms the power law model on both bacterial networks. Surprisingly, the power law model outperforms the empirical model on both eukaryote's networks.

\begin{table}[tbhp]
\caption{Prediction accuracy for the impact degree distributions: Kolmogorov-Smirnov (KS) distance, defined as $\sup_{x}|R(x)-M(x)|$, where $R(x)$ and $M(x)$ are empirical distributions and theoretical distributions, respectively.
The parenthetic values indicate the logarithmic $P$-values $p$ from the KS test, defined as $-\log_{10}(p)$.
The emphasized values correspond to the best accuracy.}
\label{table:accuracy}
\begin{center}
\begin{tabular}{l|ccc}
\hline
\hline
Species & Poisson model & Power-law model & Empirical model\\
\hline
{\it Escherichia coli} & 0.11 (2.06) & 0.08 (0.86) & {\bf 0.05 (0.27)}\\
{\it Bacillus subtilis} & 0.18 (4.65) & 0.08 (0.75) & {\bf 0.05 (0.15)}\\
{\it Saccharomyces cerevisiae} & 0.14 (2.76) & {\bf 0.07 (0.40)} & 0.07 (0.54)  \\
{\it Homo sapiens} & 0.12 (3.12)  & {\bf 0.09 (1.86)}  & 0.14 (5.09)\\
\hline
\hline
\end{tabular}
\end{center}
\end{table}

\section{Discussion and conclusion}
We proposed a model to estimate the impact degree distributions in metabolic networks, using a branching process approximation, and demonstrated its validity on real data.

The power-law model could more accurately estimate the impact degree distributions in real metabolic networks than the Poisson model because the number of offsprings for each reaction node is assumed to follow the power-law distribution.
Especially, the power-law model showed the significant agreements between the predicted distribution and the observed distribution although the case of {\it H. sapiens} represented the small $P$-value for the KS test (i.e., the low probability that the distribution is similar between models and observed data).  

However, there is no great difference of the prediction accuracy for estimating the impact degree distributions between the power-law model and the Poisson model; thus, the Poisson model may be useful for a rough estimate of the impact degree distributions.

Intrinsically, the distribution of the total number of offsprings [i.e., $P(r)$] is not significantly different between the power-law model and the Poisson model in the case of smaller $r$.
As a simple example, we here consider the case of $\mu=1$ (i.e., the critical case).
In this case, the impact degree distributions $P(r)$ of the power-law model and the Poisson model correspond to $\propto r^{-(1+1/\gamma)}$ and $\propto r^{-3/2}$, respectively.
Especially, $P(r)$ of the power-law model is the power-law distribution with the exponent ranging between $-2$ and $-1.5$ because of $1<\gamma<2$; thus it is not critically different from $r^{-3/2}$ of the Possion model for smaller $r$.
Since the impact degrees of real metabolic networks (i.e., $r$) were relatively small ($r<60$), there might be no great difference of the prediction accuracy between these models.

The similarity of predicted distributions of the total number of offsprings between the Poisson model and the power-law model is also explained using the Otter's theorem [Theorem 13.1 in \cite{Harris1989}].
This theorem indicates that the distribution $P(r)$ of the total number of offsprings has the universal property of a power-law tail with the exponent $-3/2$ as $r \to \infty$ under mild conditions on offspring distributions, and it implies that the types of offspring distributions hardly influence the distribution of the total number of offsprings (i.e., impact degree distributions).
Note the Otter's theorem does not contradict with the analytical distribution $P(r)$ of the power-law model because this theorem is not directly applicable to the power-law model due to the different assumptions in the derivation of $P(r)$ between the power-law model and the Otter's theorem.

The prediction performance is influenced by the assumption of offspring distributions and the fidelity of branching processes.
To purely evaluate the validity of branching process approximation, the empirical model is useful because of no assumption of offspring distributions.
It is expected that the empirical model show the best prediction accuracy because of using empirical offspring distributions.
In the case of bacteria (i.e., {\it E. coli} and {\it B. subtilis}), this expectation is true, suggesting that the branching process approximation is useful for estimating the impact degree distributions.
In the case of eukaryotes (i.e., {\it S. cerevisiae} and {\it H. sapiens}), on the other hand, we observed the unexpected results: the empirical model shows the relatively-low prediction accuracy.
Especially, the prediction accuracy of the empirical model is lowest in the case of {\it H. sapiens} (human).
This result implies limitations to the estimation of impact degree distributions based on the branching process approximation in the case of metabolic networks of eukaryotes (i.e., higher organisms).

A limitation of the branching process approximation is that we need to assume tree structures of networks (i.e., no cycles).
The presence of cycles may lead to an overestimation of the number of offsprings $d_i$ for each reaction, because some offsprings of a progenitor (reaction node) may have already been inactivated due to cycle structures.
From this reason, the models may overestimate the impact degree distributions.
On the other hand, however, some reactions with more than one incoming edge may be inactivated in the presence of cycles, if all their parents are inactivated through different paths.
In this case, the number of offsprings is underestimated.
The estimation of the number of offsprings depends on the relative importance of these two effects, and it may be not simple.
The difficulty in this estimation is also a limitation of the model.

The branching process model needs to be improved by considering additional assumptions other than metabolic network structures in order to obtain better predictions.
For example, the assumption of variable propagations in the branching process \cite{Dobson2010}, in which the mean of offspring distributions differs at each propagation stage, may be useful because the degree of propagation may depend on metabolic dynamics such as gene expressions and metabolite concentrations. 
To apply this modified branching process model, time-series data on metabolic dynamics after gene disruptions, which are obtained by metabolomic analysis, are necessary for estimating the mean of offspring distribution at each propagation stage (i.e., time after gene disruptions).
Since such data are unavailable at present, however, it is difficult to apply this modified branching process model.
Similarly, other biologically-suitable assumptions are hardly determined because of few observed data on metabolic dynamics.
Although there are above constraints on observed data on metabolic dynamics, we believe that the consideration of the additional information improves the prediction of impact degree distributions.
In the future, the improvement of the prediction of the impact degree distributions using the branching process approximation may be possible with available data on metabolic dynamics.

The branching process approximation is useful for estimating the impact degree distributions in metabolic networks although it has the above limitations; thus, it may be a powerful tool for evaluating a robustness of biological systems.

\section*{Acknowledgment}
This work was partially supported by the JST PRESTO program.
TA, TT and JPV were partially supported by a JSPS/INSERM grant.
TA was partially supported by Grant-in-Aid \#22650045 from MEXT, Japan.

 

\end{document}